\documentclass[iop, numberedappendix, appendixfloats]{emulateapj}
\usepackage{graphicx,natbib,amsmath,multirow,hyperref,booktabs}

\shorttitle{Solar Cycle Propagation, Memory, and Prediction}
\shortauthors{Mu\~noz-Jaramillo, Dasi-Espuig, Balmaceda \& DeLuca}

\begin{document}

\title{Solar Cycle Propagation, Memory, and Prediction: Insights from a Century of Magnetic Proxies}

\author{Andr\'es Mu\~noz-Jaramillo\altaffilmark{1, 2, 3, *}, Mar\'ia Dasi-Espuig\altaffilmark{4}, Laura A.\ Balmaceda\altaffilmark{5}, and Edward E. DeLuca\altaffilmark{1}}
\affil{$^1$ Harvard-Smithsonian Center for Astrophysics, Cambridge, MA 02138, USA; amunoz@cfa.harvard.edu}
\affil{$^2$ University Corporation for Atmospheric Research, Boulder, CO 80307, USA}
\affil{$^3$ Department of Physics \& Astronomy, University of Utah, Salt Lake City, UT 84112, USA}
\affil{$^4$ Max-Planck-Institut f\"ur Sonnensystemforschung, Katlenburg-Lindau, Germany}
\affil{$^5$ Institute for Astronomical, Terrestrial and Space Sciences (ICATE-CONICET),San Juan, Argentina}
\affil{$^*$ correspondence should be sent to: \href{mailto:amunoz@cfa.harvard.edu}{amunoz@cfa.harvard.edu}}

\begin{abstract}
The solar cycle and its associated magnetic activity are the main drivers behind changes in the interplanetary environment and the Earth's upper atmosphere (commonly referred to as space weather).  These changes have a direct impact on the lifetime of space-based assets and can create hazards to astronauts in space.  In recent years there has been an effort to develop accurate solar cycle predictions (with aims at predicting the long-term evolution of space weather), leading to nearly a hundred widely spread predictions for the amplitude of solar cycle 24.  A major contributor to the disagreement is the lack of direct long-term databases covering different components of the solar magnetic field (toroidal vs.\ poloidal). Here we use sunspot area and polar faculae measurements spanning a full century (as our toroidal and poloidal field proxies), to study solar cycle propagation, memory, and prediction.  Our results substantiate predictions based on the polar magnetic fields, whereas we find sunspot area to be uncorrelated to cycle amplitude unless multiplied by area-weighted average tilt.  This suggests that the joint assimilation of tilt and sunspot area is a better choice (with aims to cycle prediction) than sunspot area alone, and adds to the evidence in favor of active region emergence and decay as the main mechanism of poloidal field generation (i.e. the Babcock-Leighton mechanism).  Finally, by looking at the correlation between our poloidal and toroidal proxies across multiple cycles, we find solar cycle memory to be limited to only one cycle.
\end{abstract}

\keywords{Sun: dynamo --- Sun: activity --- Sun: surface magnetism}

\section{Introduction}

The solar magnetic cycle is without a doubt the main driver behind changes in the heliospheric environment (Schwenn 2006\nocite{schwenn2006b}), violent activity that shapes the Earth's magnetosphere (Pulkkinen 2007\nocite{pulkkinen2007}), and $>80\%$ of the Sun's radiative output variability (Domingo et al.\ 2009\nocite{domingo-etal2009}) --  having the potential of disrupting communications, satellites, and power distribution systems; as well as being potentially hazardous to passengers traveling in high-altitude polar routes and astronauts in space.   Because of this, there has been a continuous effort to develop accurate solar cycle predictions (with aims at predicting the long-term evolution of space weather).

Cycle predictions are typically classified into three categories: extrapolation methods, which use the mathematical properties of the sunspot data series to predict future levels of activity; precursor methods, which use different measurable quantities as a proxy to estimate the subsequent cycle's amplitude; and model-based predictions which use the assimilation of data into models of the solar cycle to make predictions (for a review on the different types of prediction see for example Petrovay 2010\nocite{petrovay2010}, and Pesnell 2012\nocite{pesnell2012}).

As opposed to extrapolation and precursor methods (which have been around for nearly 50 years), model-based predictions made their first debut at the end of solar cycle 23.  There were three different model-based predictions for cycle 24:  two predictions based on mean-field kinematic dynamos (Dikpati et al.\ 2006\nocite{dikpati-detoma-gilman2006}; Choudhuri et al.\ 2007\nocite{choudhuri-chatterjee-jiang2007}), and a prediction using a low-mode model with variable magnetic helicity which assimilates data using an ensemble Kalman filter (Kitiashvili \& Kosovichev 2008\nocite{kitiashvili-kosovichev2008}).   The application of mean-field kinematic dynamos to solar cycle predictions has attracted a lot of attention because, in spite of using similar models, dynamo-based predictions turned out quite different:  Dikpati et al.\ (2006\nocite{dikpati-detoma-gilman2006}) predicted a stronger cycle 24 than cycle 23 ($SSN = 155-180$), whereas Choudhuri et al.\ (2007\nocite{choudhuri-chatterjee-jiang2007}) predicted a weaker cycle 24 than cycle 23 ($SSN = 80$).   Nevertheless, although similar in nature, the differences between these two models are subtle but significant.  In this work we focus on two of them:
\begin{enumerate}
  \item The quantities used to make cycle predictions (sunspot area vs.\ dipolar moment),
  \item The relative importance between diffusive and advective transport of the solar magnetic fields (which sets the memory of the solar dynamo).
\end{enumerate}
One of the main obstacles preventing us from reaching consensus regarding these issues is our inability to observe the solar magnetic field inside the convection zone -- leaving surface observations as our only window into the inner workings of the dynamo.  To compound the problem, systematic magnetic field measurements are only available since the dawn of the space age (spanning only 4 cycles) making long-term magnetic proxies necessary.

\begin{figure*}
\centering
\begin{tabular}{c}
  \includegraphics[width=0.85\textwidth]{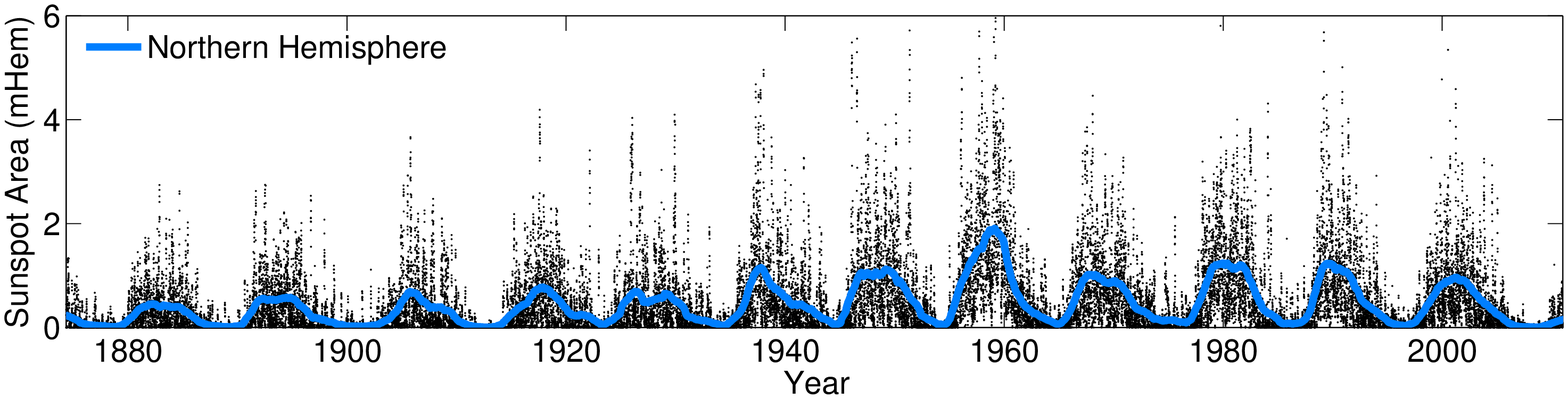}\\
  \includegraphics[width=0.85\textwidth]{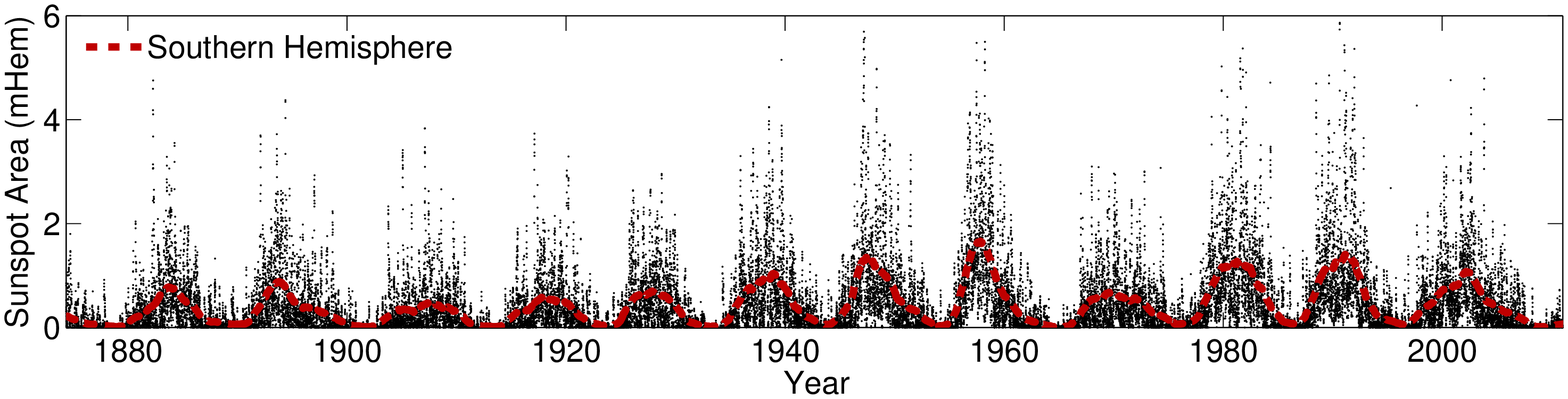}
\end{tabular}
\caption{Total daily sunspot area (black dots) is calculated for the northern (top panel) and southern (bottom panel) hemispheres.  A 24-month Gaussian filter is applied to remove the high-frequency component in the data series shown as a solid line for the northern (top panel) hemisphere and as a dashed line for the southern (bottom panel) hemisphere.}\label{Fig_Dat}
\end{figure*}

The solar cycle can be understood as a process that brings the solar magnetic field (back and forth) from a configuration that is predominantly poloidal (contained inside the $r-\theta$ plane), to one predominantly toroidal (wrapped around the axis of rotation; locally perpendicular to the $r-\theta$ plane).  From this follows that at least one poloidal and one toroidal field proxies are necessary.  Here we use a recently standardized database of polar magnetic flux measurements (based on polar faculae observations) going back to the beginning of the 20th century (Mu\~noz-Jaramillo et al.\ 2012)\nocite{munoz-etal2012}, and a long-term homogeneous sunspot area database (Balmaceda et al.\ 2009)\nocite{balmaceda-etal2009}, to validate the use of different magnetic proxies in the context of cycle prediction (Section \ref{Sec_Prediction}), and better understand the issue of solar cycle memory (Section \ref{Sec_Memory}).  Discussion and conclusions can be found in Section \ref{Sec_Conclusions}.

\section{Sunspot Area: Data and Smoothing}

In this work we use a homogeneous database of sunspot areas (Balmaceda et al.\ 2009\nocite{balmaceda-etal2009}) mainly based on observations taken by the Royal Greenwich Observatory, several stations belonging to the former USSR (compiled in the \emph{Solnechniye Danniye} bulletin issued by the Pulkovo Astronomical Observatory), and the US Air Force Solar Optical Observing Network (SOON).  We separate the data in Northern (top panel in Figure \ref{Fig_Dat}) and Southern (bottom panel in Figure \ref{Fig_Dat}) hemisphere sets, calculating the total hemispheric daily sunspot area.  Area belonging to groups observed at the equator are not assigned to any of the two hemispheres.

We remove high-frequency components by convolving our data series with the modified 24-month Gaussian filter:

\begin{equation}\label{Eq_Filter}
        \begin{array}{l}
          F(t,t',a)\\
          = \left\{\begin{array}{ccc}
                     0 & & t \leq t' - a\\
                     e^{-(t - t')^2/(2a^2)} - \frac{3 - (t - t')^2}{2a^2} e^{-2} & & t' - a < t \leq t' + a\\
                     0 & & t > t' + a
               \end{array}\right., 
        \end{array}
\end{equation}
where $t'$ denotes the position of the center and $a=12$ months the half-width of the Gaussian filter.  This type of filter has been found to yield more consistent results while finding maxima and minima (using different databases like the international sunspot number and the 10.7cm radio flux), than the traditional 13-month running mean (Hathaway 2010)\nocite{hathaway2010}.

\begin{figure*}
\centering
\begin{tabular}{c}
  \includegraphics[width=.7\textwidth]{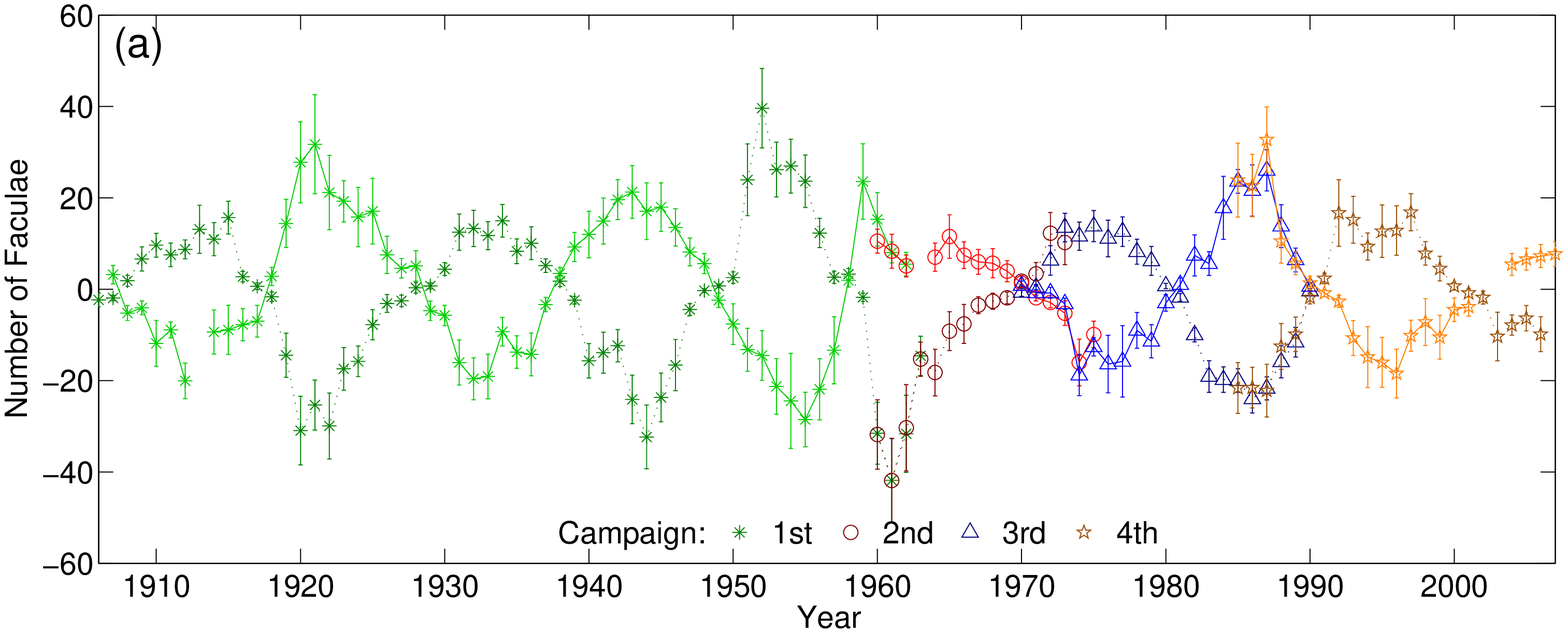}\\
  \includegraphics[width=.7\textwidth]{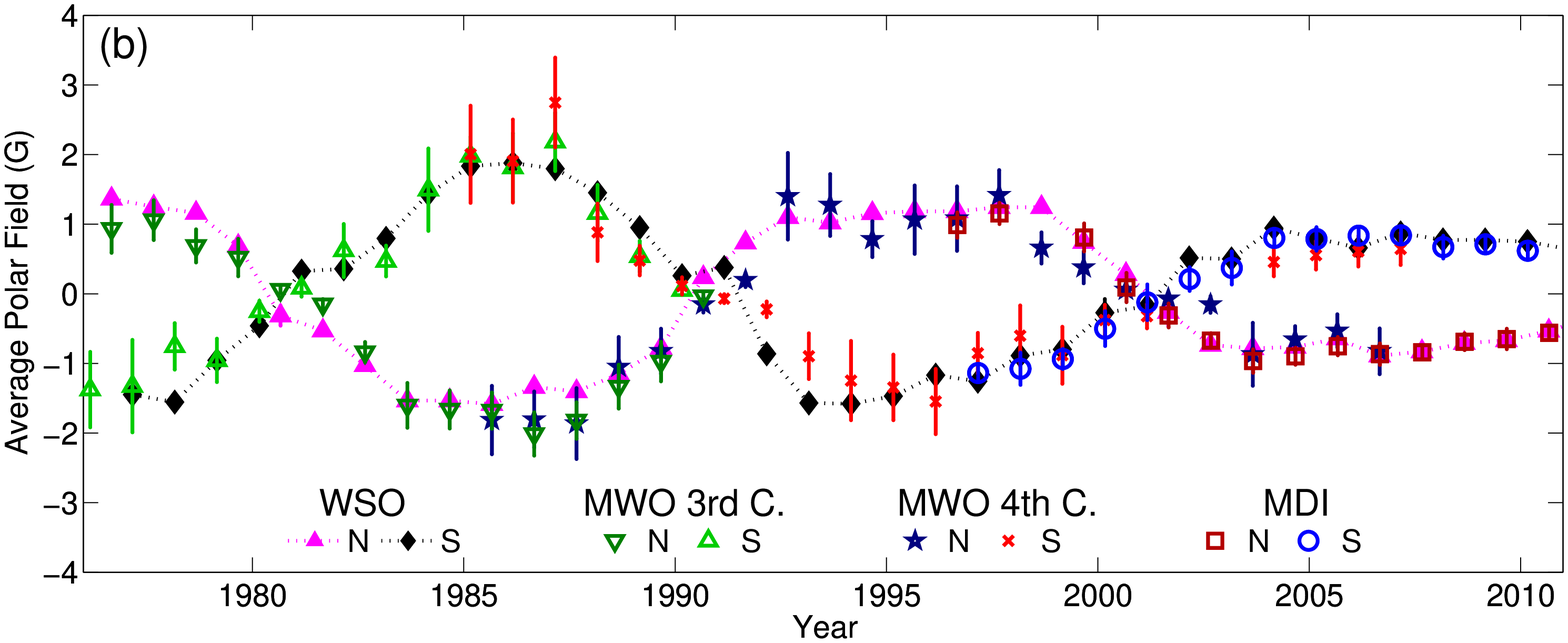}\\
  \includegraphics[width=0.76\textwidth]{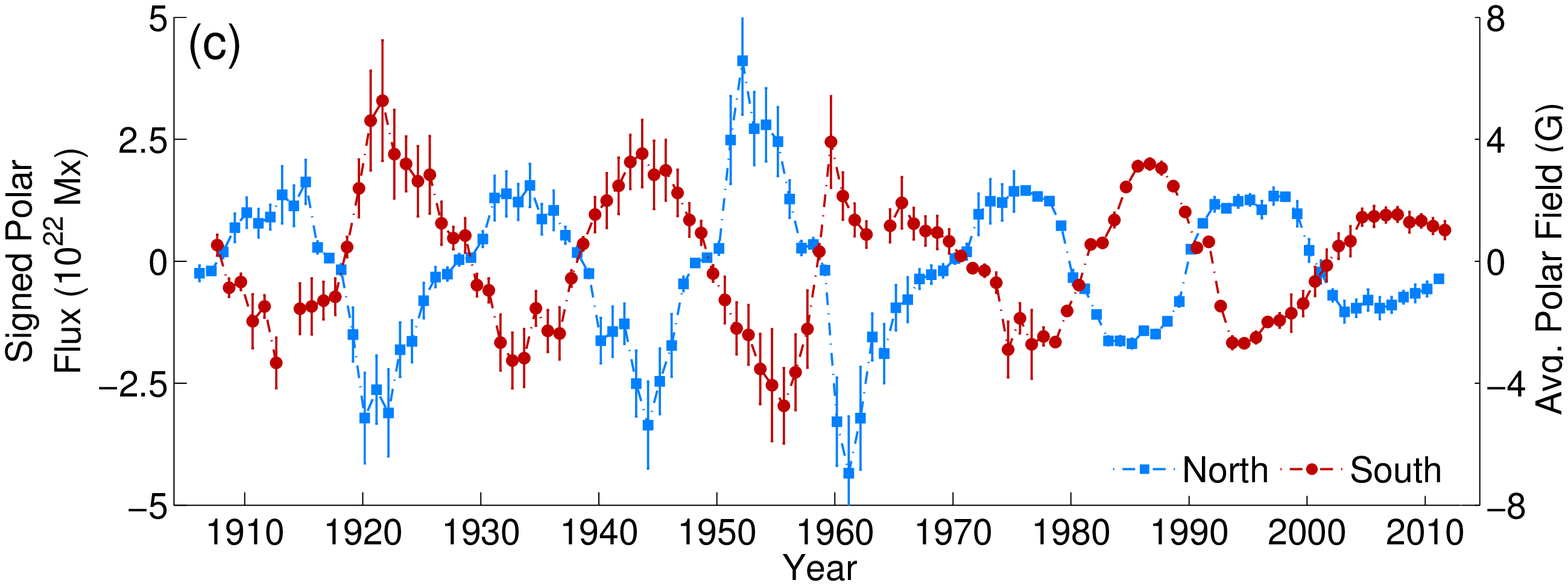}
\end{tabular}
\caption{Four Mount Wilson Observatory campaigns are standardized using their overlap to obtain a consistent polar faculae database (a).  Different colors and markers correspond to different data reduction campaigns.  This database is calibrated using data from the Wilcox Solar Observatory and the MDI, in order into convert it to polar flux estimates (b).  The resultant databases are consolidated into a single proxy (c).}\label{Fig_PF_Dat}
\end{figure*}

\section{Polar Faculae as a Proxy for Polar Magnetic Flux}

Our polar flux database comes from a recent calibration and standardization (Mu\~noz-Jaramillo et al.\ 2012\nocite{munoz-etal2012}) of four Mount Wilson Observatory data reduction campaigns (Sheeley 1966, 1976, 1991, 2008\nocite{sheeley1966,sheeley1976,sheeley1991,sheeley2008}).   Consecutive campaigns were cross-calibrated using five year overlaps between different data reduction campaigns (see Figure \ref{Fig_PF_Dat}-a).  The cross-calibrated dataset was validated using an automatic detection algorithm on intensity data from the Michelson Doppler Imager (MDI; Scherrer et al.\ 1995\nocite{scherrer-etal1995}) on board the SOlar and Heliospheric Observatory (SOHO) spacecraft.  The resultant faculae database was calibrated in terms of polar magnetic field and flux using magnetic field measurements taken by the Wilcox Solar Observatory and SOHO/MDI (see Figure \ref{Fig_PF_Dat}-b).   Our results demonstrate that there is a strictly proportional relationship between facular count, average polar field, and total signed polar flux (during all phases of the cycle and for each hemisphere separately) making it an ideal proxy for the evolution of the polar magnetic fields. Once converted into polar flux values we combine MWO, WSO and MDI data into a single consolidated database (see Figure \ref{Fig_PF_Dat}-c) and use average polar flux during solar minimum as a proxy for the poloidal component of the magnetic field  (for more details on the validation and magnetic calibration of polar faculae data please refer to Mu\~noz-Jaramillo et al.\ 2012\nocite{munoz-etal2012}).

\section{Relationship Between the Poloidal and Toroidal Field Proxies:  Implications for Current Model-Based Predictions}\label{Sec_Prediction}

While there is consensus on the process which converts poloidal to toroidal field (stretching of poloidal field by differential rotation; Parker 1955a\nocite{parker1955a}), there are several mechanisms which may be playing a role in the conversion from toroidal to poloidal field.  Chief among them are: the twisting of toroidal field into the poloidal plane due to its interaction with helical turbulent convection (Parker 1955a\nocite{parker1955a}), tilted active region (AR) emergence and decay (also known as the Babcock-Leighton mechanism, Babcock 1961; Leighton 1969\nocite{babcock1961,leighton1969}), flux-tube instabilities (Schmitt 1987\nocite{schmitt1987}), and hydrodynamical shear instabilities (Dikpati \& Gilman 2001\nocite{dikpati-gilman2001}) -- for a comprehensive review please refer to Charbonneau (2010\nocite{charbonneau2010}) and references therein.

Currently, the main contending theory for the generation of poloidal field is the Babcock-Leighton (BL) mechanism.  Supporting evidence comes in the form of surface flux-transport simulations showing that AR emergence and decay leads to the reversal and concentration of polar flux (Wang, Nash \& Sheeley 1989\nocite{wang-nash-sheeley1989b}), and which driven by semi-synthetic records of sunspot groups obtain a significant correlation between the polar field at minimum and the strength of the subsequent sunspot cycle (Jiang et al.\ 2011)\nocite{jiang-etal2011}.  Furthermore, a significant correlation has been found between AR tilt and the amplitude of the next cycle (Dasi-Espuig et al.\ 2010\nocite{dasiespuig-etal2010}), suggesting a direct connection between AR properties and cycle propagation.

\begin{figure}
\centering
  \includegraphics[width=.4\textwidth]{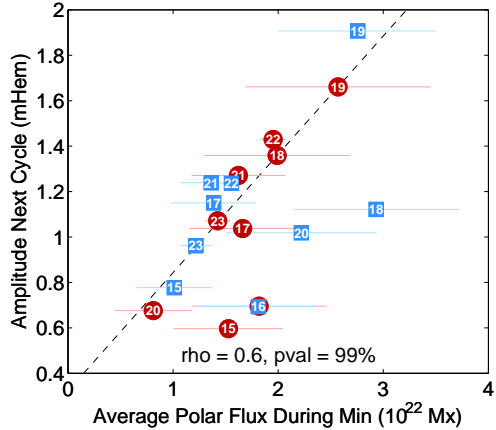}
\caption{Correlation between polar flux at solar minimum and the amplitude of the next cycle. Square (circular) markers denote data for the northern (souther) hemispheres.  Markers are numbered using cycle amplitude as reference.  The dashed line corresponds to a linear fit using the least absolute residuals method.  The text inside the figure panel indicates the Pearson's correlation coefficient and its statistical significance.}\label{Fig_Pol_tor}
\end{figure}

From a practical point of view, the most attractive feature of the BL mechanism is the fact that ARs (and their role in the evolution of the solar polar magnetic field) play a crucial role in the progression of the cycle, allowing modelers to use surface magnetic field observations to drive dynamo-based predictions.   In particular, Dikpati et al.\ (2006\nocite{dikpati-detoma-gilman2006}) used sunspot area to drive the generation of poloidal field in their model-based prediction, whereas Choudhuri et al.\ (2007\nocite{choudhuri-chatterjee-jiang2007}) bypassed the generation of poloidal field by directly using the axial dipole moment component of the solar magnetic field calculated from the polar fields by Svalgaard et al.\ (2005\nocite{svalgaard-etal2005}).  The axial dipole moment at minimum has been found to be a good predictor for cycle amplitude; both when calculated using direct measurements of the polar fields, for cycles 22-24 (Schatten 2005\nocite{schatten2005}; Svalgaard et al.\ 2005\nocite{svalgaard-etal2005}), and when estimated using solar open magnetic flux derived from the historical \emph{aa} index (for cycles 12-23), after removing the contribution of the solar wind speed (Wang \& Sheeley 2009\nocite{wang-sheeley2009}).  Taking advantage of our century of toroidal and poloidal proxies, our first task is to look at their relationship from the point of view of cycle prediction.

We find a good correlation between polar flux at minimum and the amplitude of the next cycle (Figure \ref{Fig_Pol_tor}), with a Pearson's correlation coefficient of $\rho = 0.60$ and $P=99\%$ confidence level.  An interesting feature of the correlation between polar flux at minimum and the next cycle's amplitude is the apparent existence of two separate branches in their relationship.  This feature becomes more evident after performing a linear fit using the least absolute residuals method (shown as a dashed line in Figure \ref{Fig_Pol_tor}) which naturally gives less weight to possible outliers in the dataset.  The separation of data into two branches results in a large improvement in the Pearson's correlation coefficient: $\rho = 0.96$ ($\rho = 0.95$) and $P=99\%$ ($P=99\%$) confidence level for the main (secondary) branches.  This suggests that finding a way to evaluate on which branch will a cycle fall would result in a highly effective method of prediction.   An in-depth study of this separation (which seems to be related to the relative strength of the dipolar and quadrupolar moments during minimum) and its application to solar cycle prediction can be found in Mu\~noz-Jaramillo et al.\ (2013\nocite{munoz-etal2013b}).

\begin{figure*}
\begin{center}
\begin{tabular}{cc}
  \includegraphics[width=0.35\textwidth]{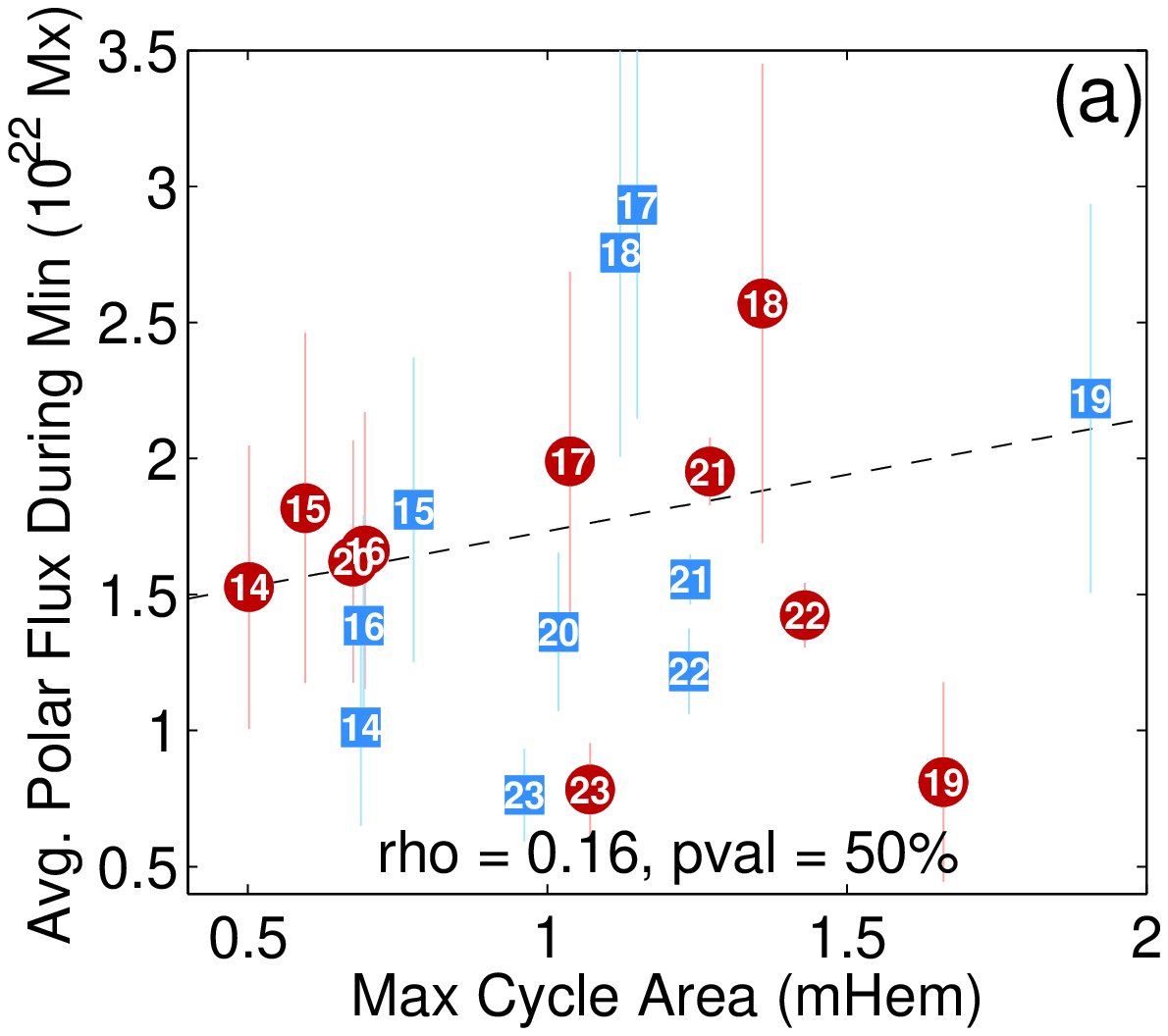} & \includegraphics[width=0.35\textwidth]{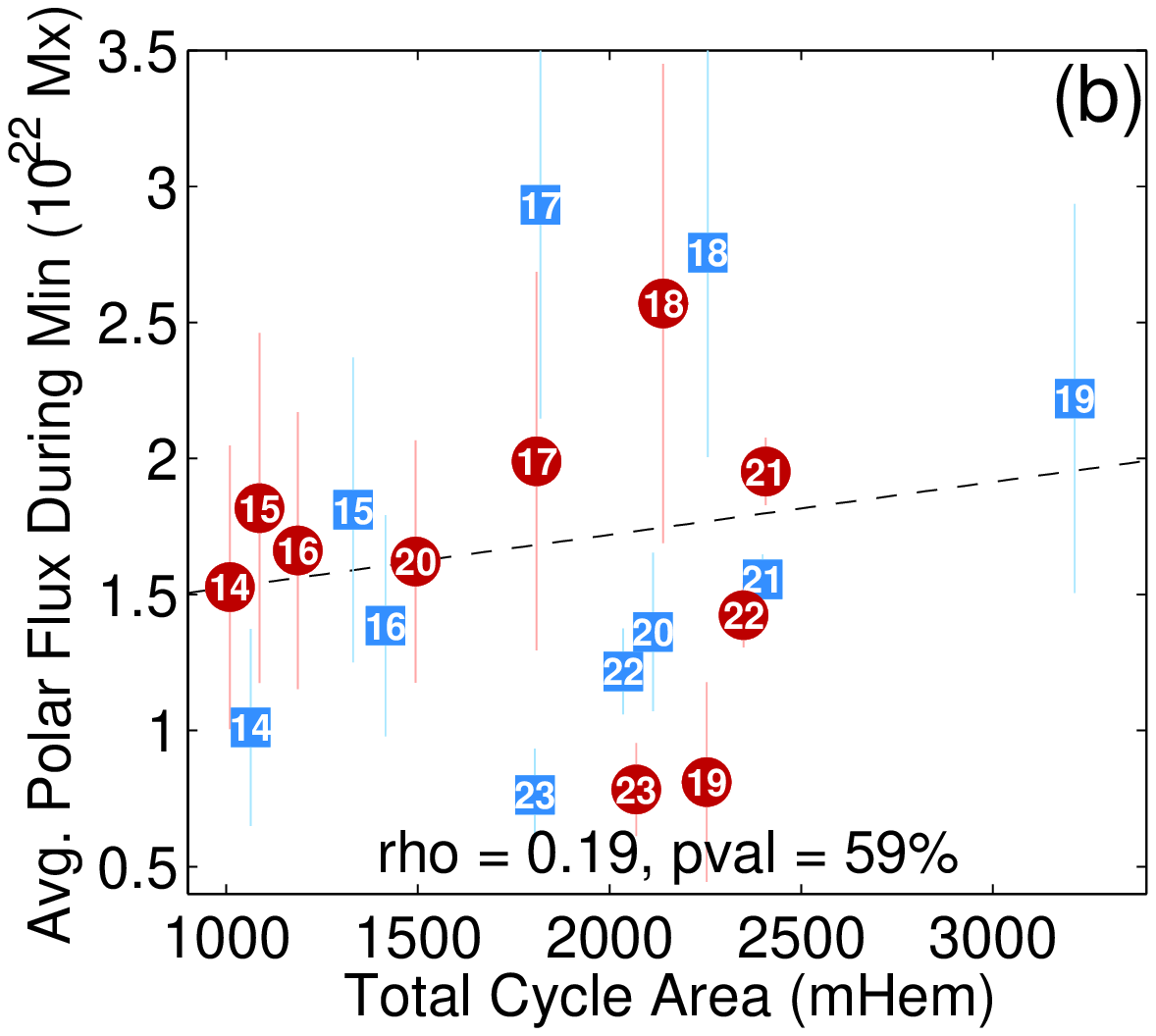}\\
  \includegraphics[width=0.35\textwidth]{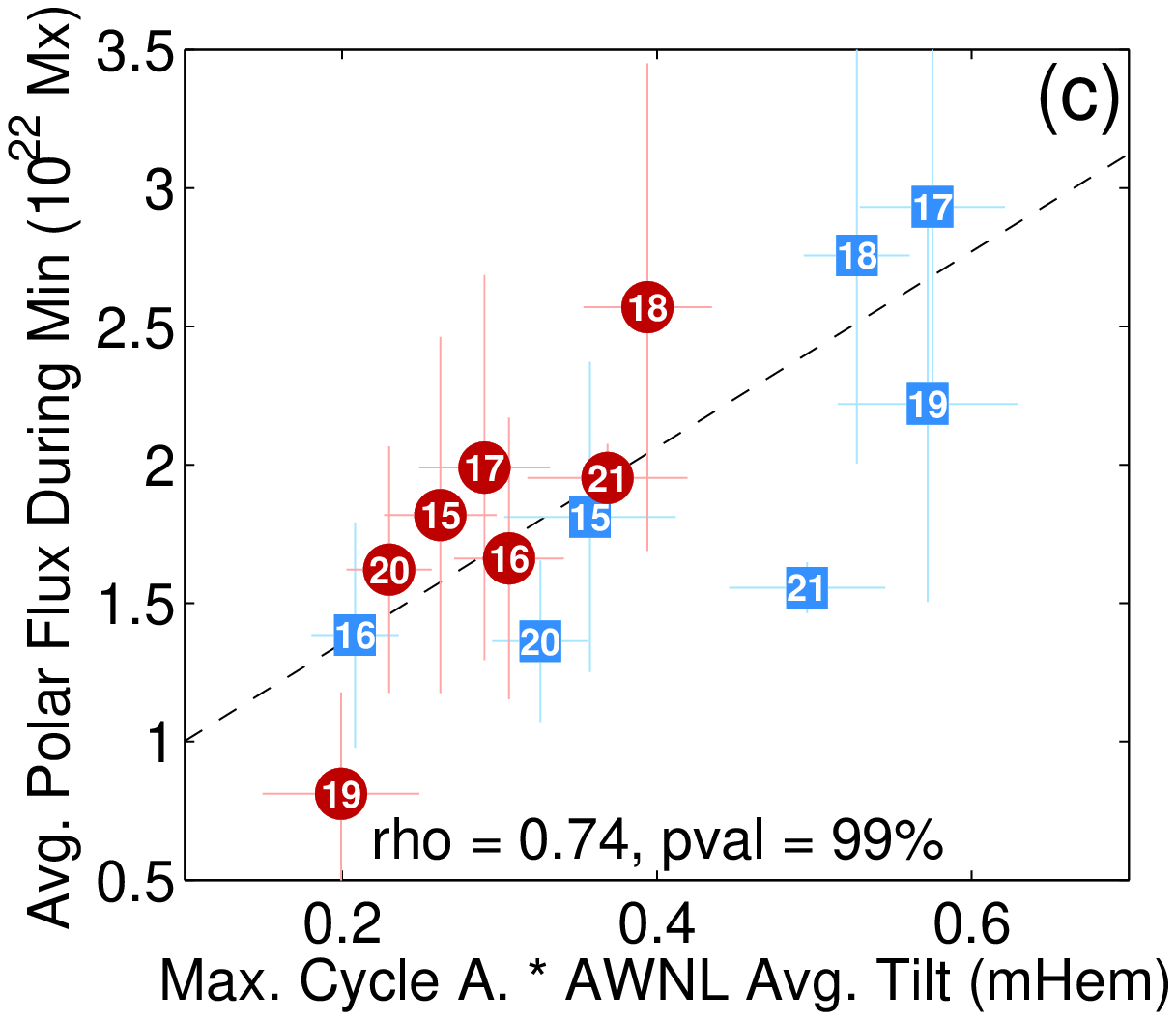} & \includegraphics[width=0.35\textwidth]{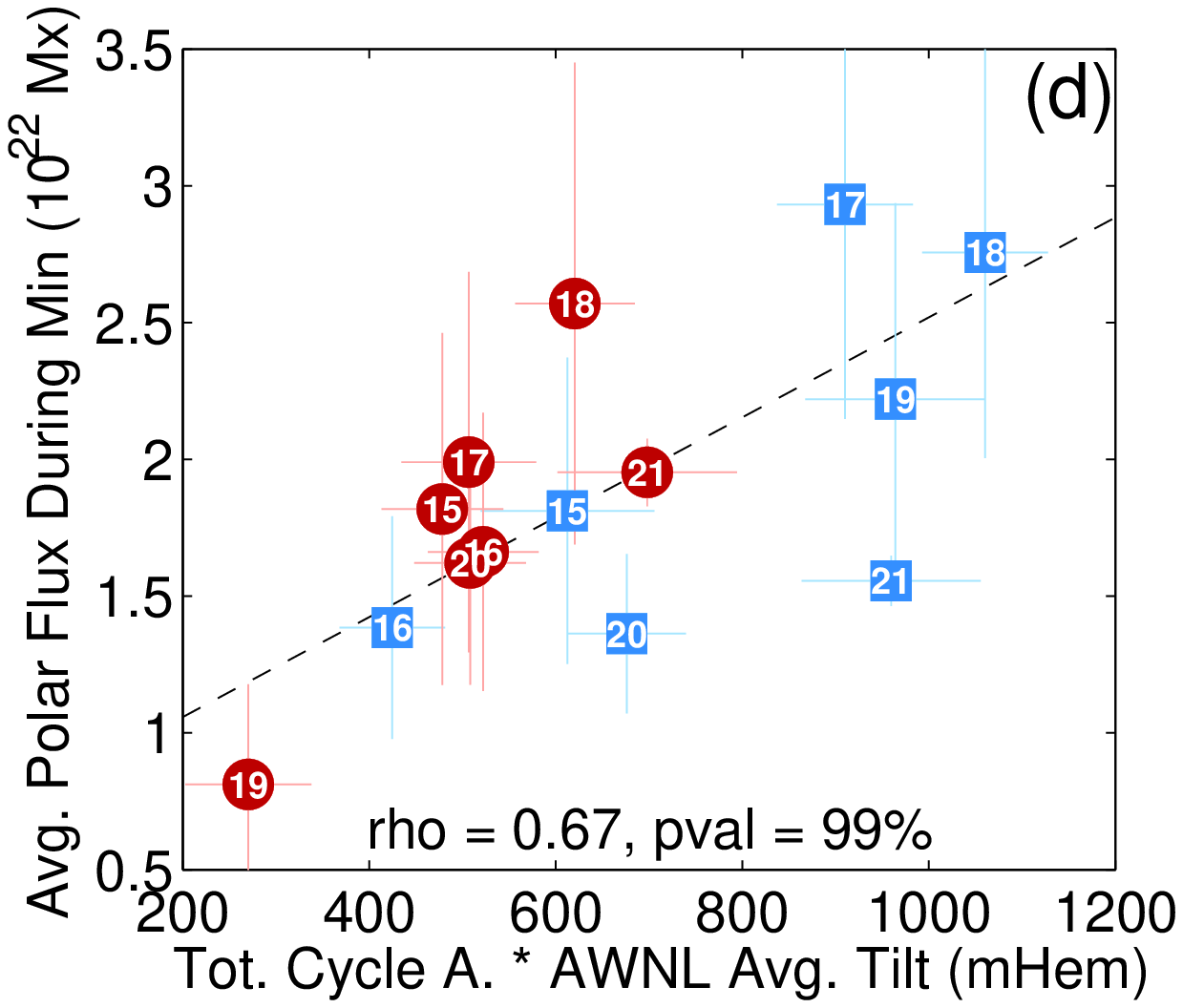}
\end{tabular}
\end{center}
\caption{We find no correlation between the maximum (a), or total (b), sunspot area of a cycle and polar flux during the subsequent minimum.  However, both maximum (c) and total (d) sunspot area become correlated with polar flux if multiplied by the area-weighted average tilt normalized by latitude of emergence (AWNL).  Square (circular) markers denote data for the northern (souther) hemispheres.  Markers are numbered using cycle amplitude as reference.  The dashed line corresponds to a linear fit using the least absolute residuals method.  The text inside figure panels indicate the Pearson's correlation coefficients and their statistical significance.}\label{Fig_Tor_pol}
\end{figure*}

Looking at the relationship between our toroidal and poloidal field proxies we find no correlation between maximum cycle sunspot area (i.e.\ cycle amplitude) and polar flux ($\rho = 0.16$ \& $P=50\%$; Fig.~\ref{Fig_Tor_pol}-a) nor between total cycle sunspot area and polar flux ($\rho = 0.19$ \& $P=59\%$;Fig.~\ref{Fig_Tor_pol}-b).   Although the apparent disconnection between sunspot area and polar flux at minimum could be interpreted as evidence against the BL mechanism, it is important to highlight that the systematic tilt presented by ARs (Hale et al.\ 1919)\nocite{hale-etal1919} is a crucial component of this mechanism of poloidal field generation.  Indeed, as shown in Figures \ref{Fig_Tor_pol}-c \& d, the multiplication of maximum cycle sunspot area, and total cycle sunspot area, by the area-weighted average tilt (normalized using latitude of emergence and calculated from Mount Wilson Observatory data; see Dasi-Espuig et al.\ 2010) turns them into quantities which are correlated to polar flux at solar minimum (with $\rho = 0.74$ \& $P=99\%$ and $\rho = 0.67$ \& $P=99\%$ respectively).   This agrees with the results of Kitchatinov \& Olemskoy (2011\nocite{kitchatinov-olemskoy2011}) who find, for cycles 19-21, a good correlation between the total contribution of all sunspot groups to the large scale dipolar field (calculated using the area of the largest spot, the angular extent, and the tilt of each group) and the amplitude of the next cycle.

These results suggests that sunspot area alone may not be an appropriate quantity for use in model-based predictions and explains partly why the models of Dikpati et al.\ (2006\nocite{dikpati-detoma-gilman2006}) and Choudhuri et al.\ (2007\nocite{choudhuri-chatterjee-jiang2007}) yielded such different predictions.  However, the improvement introduced by including tilt suggests that future model-based predictions which rely on AR data will likely obtain better performance if they assimilate tilt, as well as sunspot area data.  It is important to note that these results are obtained using a simplified way of including information on the spatio-temporal distribution, time and latitude of emergence, of ARs (as performed by Dasi-Espuig et al.\ 2010).  A detailed assimilation of this information (only possible using a more sophisticated model) leads to a significant improvement in the estimation of solar minimum conditions (see Cameron et al.\ 2010\nocite{cameron-etal2010} and Cameron \& Sch\"ussler 2012\nocite{cameron-Schussler2012}), and will likely be an integral component of future model-based predictions.

\section{Observational Study of Solar Cycle Memory}\label{Sec_Memory}

The difference between the model-based predictions of Dikpati et al.\ (2006\nocite{dikpati-detoma-gilman2006}) and Choudhuri et al.\ (2007\nocite{choudhuri-chatterjee-jiang2007}) that has received the greatest amount of attention is the relative importance of advective and diffusive transport.   On the one hand, Dikpati et al.\ (2006\nocite{dikpati-detoma-gilman2006}) use a low-diffusivity model in which the meridional flow -- a poleward plasma flow observed in near surface layers (Komm et al.\ 1993\nocite{komm-howard-harvey1993}; ) which is believed to turn around into an equatorward flow near the bottom of the convection zone, driving the equatorial migration of active latitudes (Choudhuri et al.\ 1995\nocite{choudhuri-schussler-dikpati1995}) -- is the most important mechanism of magnetic flux-transport.   On the other hand, Choudhuri et al.\ (2007\nocite{choudhuri-chatterjee-jiang2007}) use a high-diffusivity model in which diffusion is the most important mechanism of magnetic flux-transport.

In an influential theoretical study of advection-dominated (AD) vs.\ diffusion-dominated (DD) model-based predictions, Yeates et al.\ (2008\nocite{yeates-nandy-mackay2008}) found that the main difference between AD and DD predictions is the memory-span of the solar dynamo.  To reach this conclusion they performed simulations in which the source of poloidal field varied stochastically with time, and looked at the correlation between polar flux at the minimum of cycle $n$ and the amplitude of cycle $n$, $n+1$, $n+2$, and $n+3$.   They found that in the DD regime polar flux at minimum is only correlated with the amplitude of the following cycle ($n+1$), whereas in the AD regime polar flux at the minimum of cycle $n$ is correlated with the amplitudes of cycle $n$, $n+1$, and $n+2$.  In a recent revision of this work including the effect of turbulent pumping -- magnetic transport associated with the morphological asymmetry between convective upflows and downflows (Tobias et al.\ 2001\nocite{tobias-etal2001}) -- Karak \& Nandy (2012\nocite{karak-nandy2012}) found that the addition of turbulent pumping removes any long-term solar cycle memory, turning AD and DD predictions undistinguishable.  Taking advantage of our long-term poloidal and toroidal proxies, we study cycle memory from an observational point of view.

\begin{figure*}
\begin{center}
\begin{tabular}{cc}
  \includegraphics[width=0.35\textwidth]{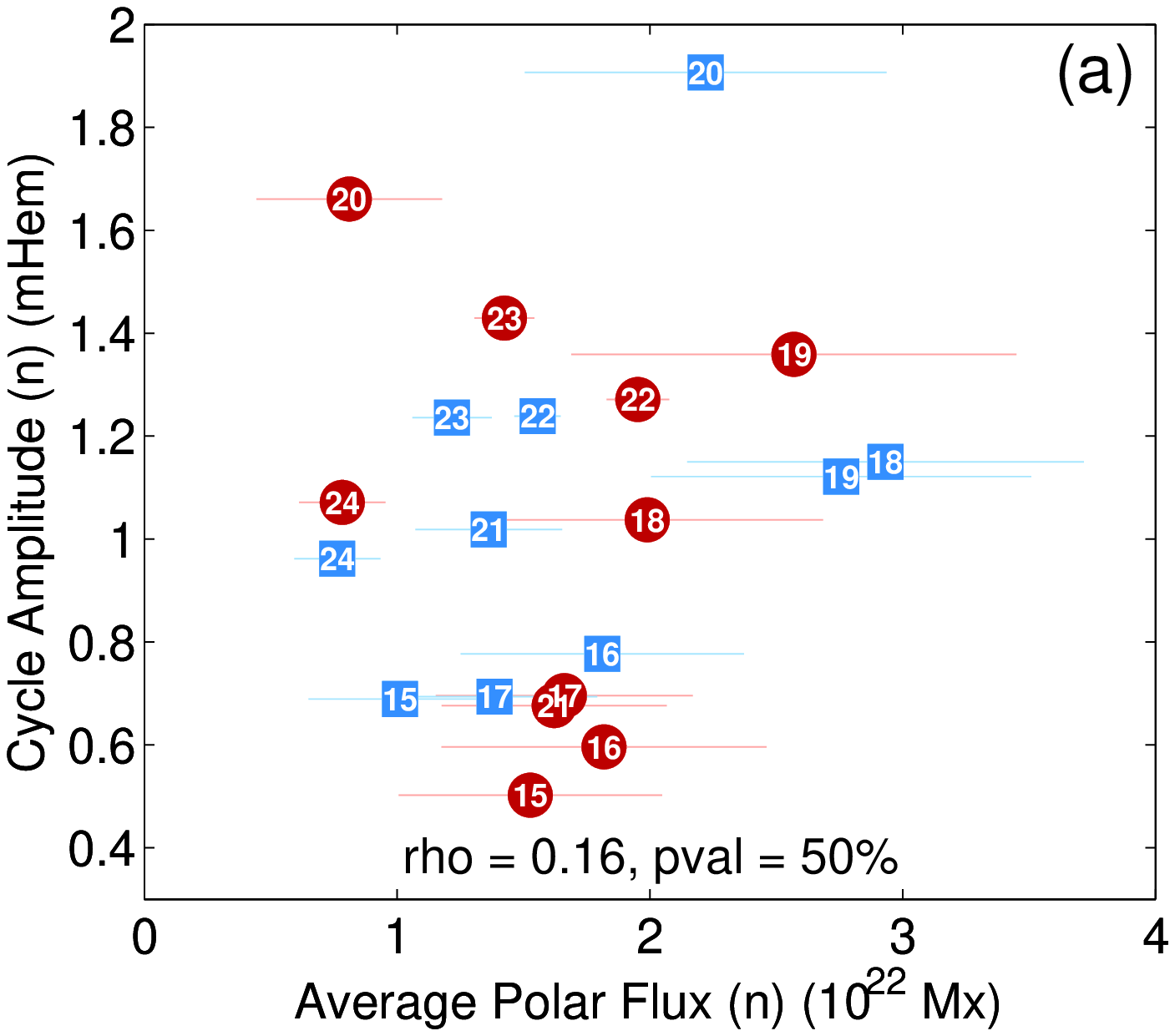} & \includegraphics[width=0.35\textwidth]{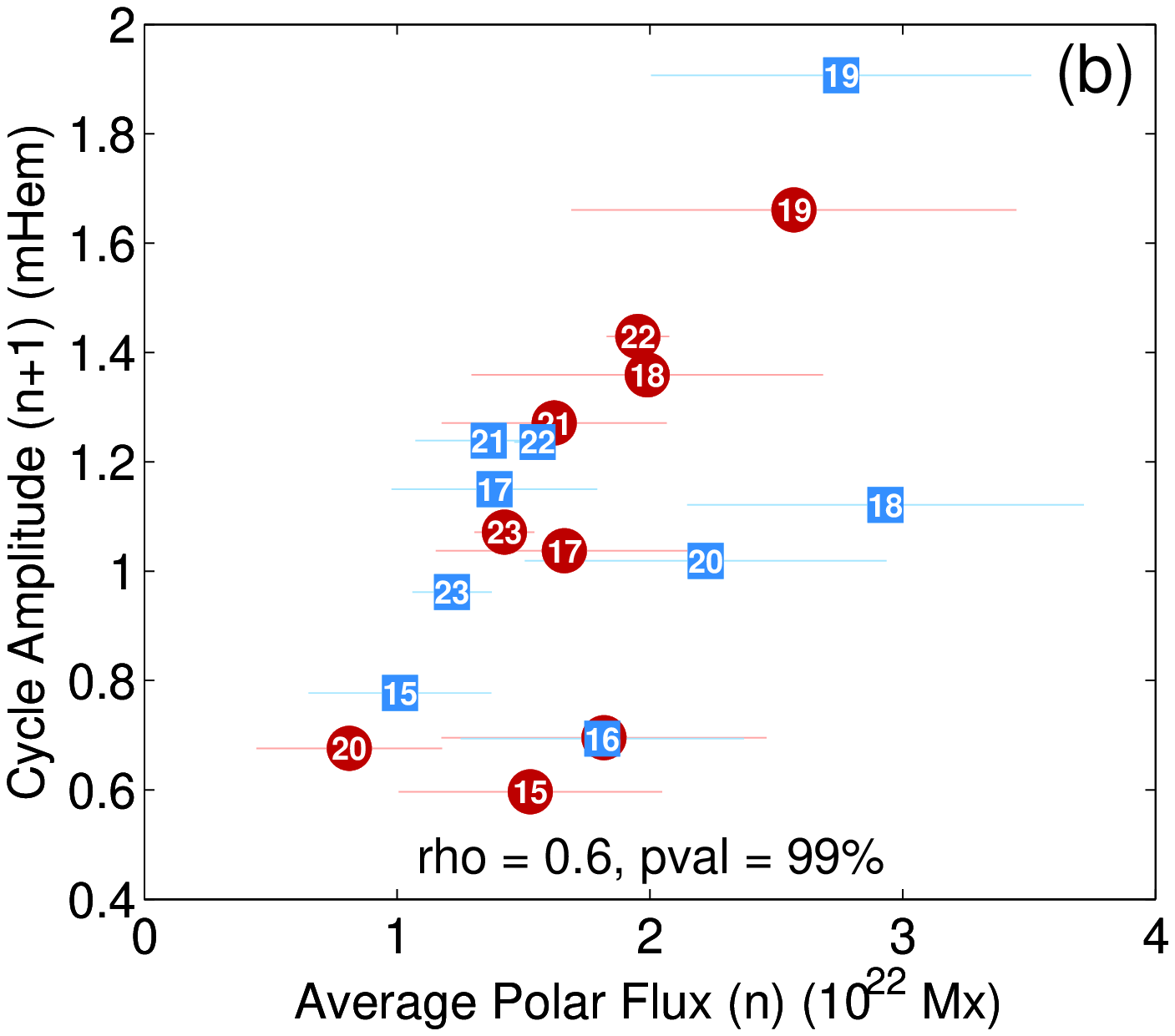}\\
  \includegraphics[width=0.35\textwidth]{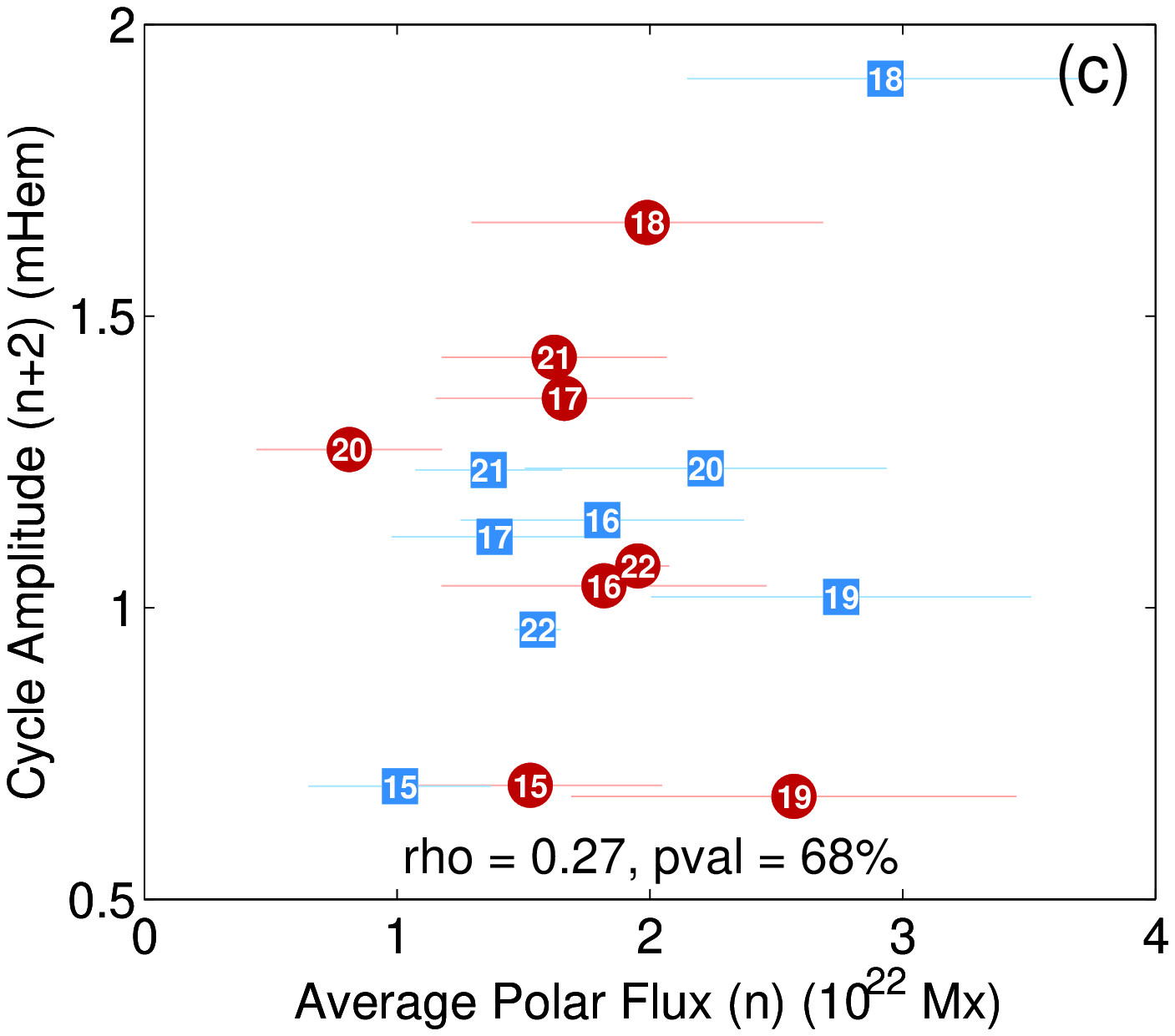} & \includegraphics[width=0.35\textwidth]{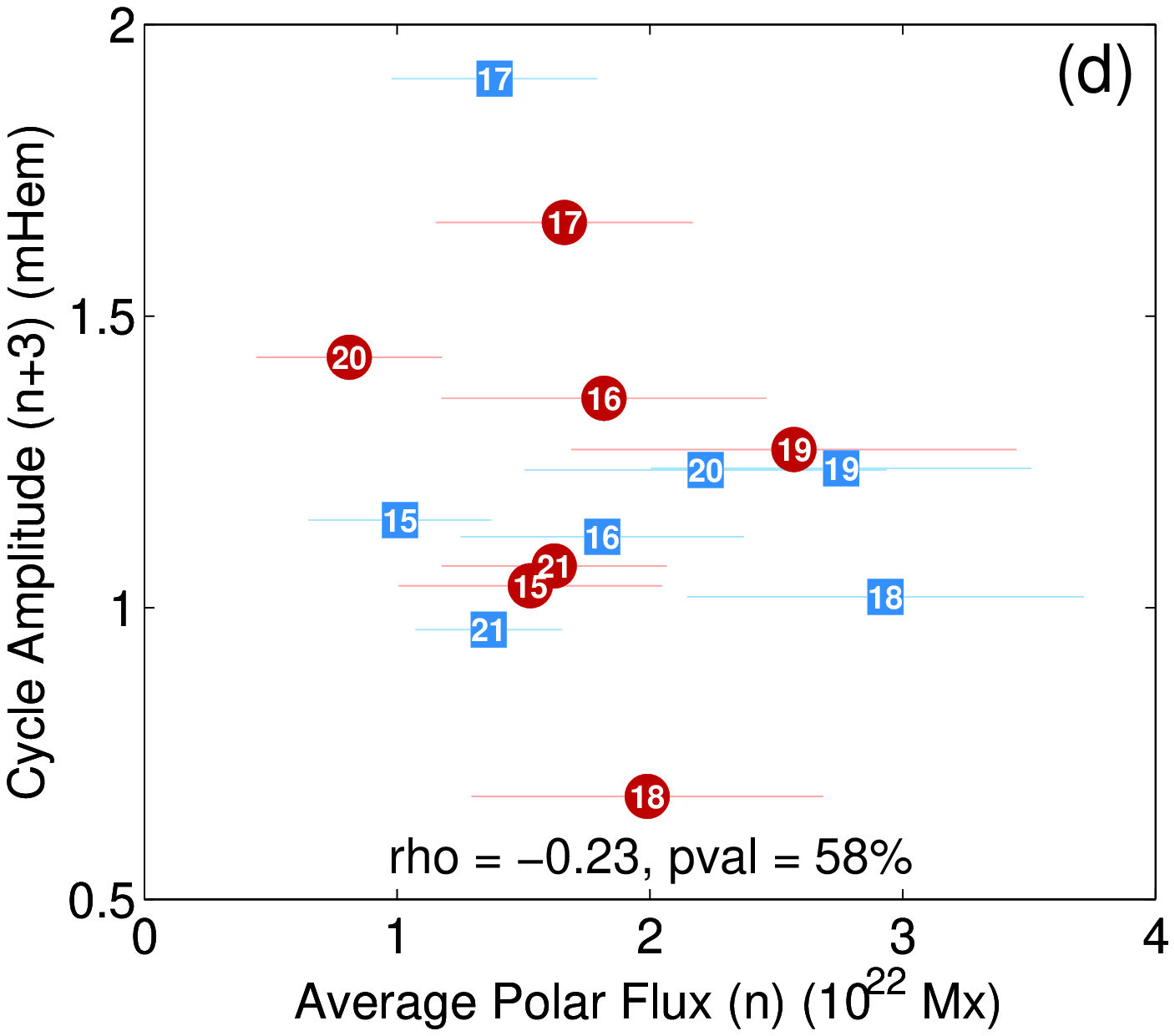}
\end{tabular}
\end{center}
\caption{Correlation between polar flux at the minimum of cycle $n$ and the amplitude of cycle $n$ (a), $n+1$ (b), $n+2$ (c), and $n+3$ (d).  Polar flux at minimum is only correlated to the amplitude of the next cycle ($n+1$; b).  This set of correlations is only consistent with a solar dynamo in which turbulent effects (diffusion and/or pumping) are as important as flux-transport by meridional flow.  Square (circular) markers denote data for the northern (souther) hemispheres.  Markers are numbered using cycle amplitude as reference. The text inside figure panels indicate the Pearson's correlation coefficients and their statistical significance.}\label{Fig_Memory}
\end{figure*}

Figure \ref{Fig_Memory} displays the correlation of polar flux at the minimum of cycle $n$ and the amplitude of cycles $n$ (a), $n+1$ (b), $n+2$ (c), and $n+3$ (d), showing only significant correlation between polar flux at the minimum of cycle $n$ and the amplitude of the next cycle ($n+1$, Fig.~\ref{Fig_Memory}-b).  In the light of the theoretical studies of Yeates et al.\ (2008\nocite{yeates-nandy-mackay2008}) and Karak \& Nandy (2012\nocite{karak-nandy2012}), these results suggest that purely advection-dominated models of the solar cycle, in which meridional flow is more important than turbulent mechanisms of flux-transport (i.e.\ turbulent diffusivity and pumping), are inconsistent with observations.  However, it is possible for random variations in meridional flow amplitude to also reduce cycle memory (removing this inconsistency).

\section{Discussion and Concluding Remarks}\label{Sec_Conclusions}

In a recent theoretical study of the dynamo basis of precursor predictions, Charbonneau \& Barlet (2011\nocite{charbonneau-barlet2011}) found that the polar field has precursor value (i.e.\ it is well correlated with the next cycle's amplitude) only when there is a connection between the surface and interior layers.  The correlation we find between polar flux at minimum and the amplitude of the next cycle demonstrates the need for such a connection between the surface magnetic field and the bottom of the convection zone.  Although high-resolution observations of the polar field show concentrated patches of magnetic field strong enough to be buoyant (Shiota et al.\ 2012\nocite{shiota-etal2012}), instead of a diffuse large-scale unipolar field, our results suggest that the total polar magnetic flux can still be seen as the surface manifestation of low-order moments in the multi-pole expansion of the solar magnetic field.   Perhaps, rather than direct subduction of polar magnetic fields, flux-transport mechanisms act on the roots of polar flux-tubes, pinning and stretching them across the bottom of the convection zone to form a large-scale poloidal field from which toroidal field is inducted.

The correlations we find between the toroidal and poloidal proxies (Figures \ref{Fig_Pol_tor} \& \ref{Fig_Tor_pol}-c \& d) represent strong observational evidence in favor of the BL mechanism as the main source of poloidal field in the Sun -- by linking together the determination of polar fields by active region emergence and decay, and its subsequent shearing by differential rotation to set the amplitude of the next cycle.  Together they support the current solar cycle logic where the two components of the solar magnetic field (toroidal and poloidal) generate each other sequentially (nicely illustrated in Charbonneau 2010\nocite{charbonneau2010}).  This has important consequences from the point of view of cycle prediction since it links the entire cycle causality to observable quantities in the photosphere.  In particular, our results substantiate proposed precursor methods based on the polar magnetic field at solar minimum (by extending their applicability to the last century) and suggests that they will be the most successful predictions once we pass solar maximum.

Taking advantage of a century of poloidal and toroidal field proxies we analyzed two of the main differences between the dynamo-based predictions of Dikpati et al.\ (2006\nocite{dikpati-detoma-gilman2006}) and Choudhuri et al.\ (2007\nocite{choudhuri-chatterjee-jiang2007}) from an observational point of view.   In the case of data assimilation we find that sunspot area is not well correlated with the amplitude of the next cycle, unless it is multiplied by area-weighted average tilt (see Fig.~\ref{Fig_Tor_pol}).  This suggests that tilt, which is a crucial component of the Babcock-Leighton mechanism, needs to be used (together with sunspot area or active region flux) in order to obtain more accurate model-based predictions.

In the case of solar cycle memory we find observations to be consistent with a short term memory (also found by Solanki et al.\ 2002\nocite{solanki-etal2002}).  This result is only naturally consistent with diffusion-dominated dynamos, or dynamos with significant transport by turbulent pumping (as proposed by Guerrero \& de Gouveia Dal Pino, 2008\nocite{guerrero-deouveiadalpino2008}).  However, a purely advection-dominated dynamo driven by a randomly fluctuating meridional flow may be able to reduce cycle memory in accordance with observations.  A detailed study of cycle memory in advection-dominated dynamos is necessary in order to address this discrepancy.

\acknowledgements

\section{Acknowledgements}

We want to thank Piet Martens and an anonymous referee for useful discussions and suggestions. This research is supported by the NASA Living With a Star Jack Eddy Postdoctoral Fellowship Program, administered by the UCAR Visiting Scientist Programs.  Andr\'es Mu\~noz-Jaramillo is very grateful to David Kieda for his support and sponsorship at the University of Utah.  E.\ DeLuca was supported by contract SP02H1701R from Lockheed Martin to SAO.

\bibliographystyle{apj}

\end{document}